\documentstyle[twocolumn,epsfig]{mn}
\begin{document}

\title{Magneto-thermal condensation modes including the effects of charged  dust particles}

\author[M. Shadmehri \& S. Dib]{Mohsen Shadmehri$^{1}$\thanks{E-mail:
mshadmehri@thphys.nuim.ie (MS); }, Sami Dib$^{2,3}$ \thanks{E-mail:sami.dib@cea.fr}\\
$^{1}$ Department of Mathematical Physics, National University Ireland,Co Kildare, Maynooth, Ireland\\
$^{2}$ Service d'Astrophysique, CEA/DSM/DAPNIA/SAp, CEA/Saclay, 91191 Gif-sur-Yvette Cedex, France\\
$^{3}$ Department of Physics, Faculty of Sciences, Lebanese University, El Hadath, Beirut, Lebanon}

\maketitle

\date{Received ______________ / Accepted _________________ }

\begin{abstract}
We study thermal instability in a magnetized and partially ionized plasma with charged dust particles.
Our linear analysis shows that the growth rate of the unstable modes in the presence of dust particles strongly depends on the ratio of the cooling rate and the modified dust-cyclotron frequency. If the cooling rate is less than the modified dust-cyclotron frequency, then growth rate of the condensation modes does not modify due to the existence of the charged dust particles. But when the cooling rate is greater than (or comparable to) the modified dust-cyclotron frequency, the growth rate of unstable modes increases because of the dust particles. Also, wavenumber of the perturbations corresponding to the maximum growth rate shifts to the smaller values (larger wavelengths) as the cooling rate becomes larger than the modified dust-cyclotron frequency. We show that growth rate of the condensation modes increases with the electrical charge of the dust particles.
\end{abstract}

\begin{keywords}
instabilities - stars: formation - ISM: general
\end{keywords}
\section{Introduction}
\label{sec:1}

One of the most important dynamical processes in astrophysical plasmas is thermal instability. A detailed analysis of thermal instability in the linear regime was given in a well-know paper by Field (1965), and in large number of subsequent works (e.g., Heyvaerts  1974; Balbus 1986; Balbus \& Soker 1989; Ib$\rm \acute{a}\tilde{n}$ez \& Escalona 1993; Steele \& Ib$\rm \acute{a}\tilde{n}$ez 1999; Burkert \& Lin 2000; Dib, Burkert \& Hujeirat 2004;  Hennebelle \& Audit 2007). The form of net cooling function is determined by the specific mechanisms of heating and cooling and it also depends on the degree of ionization of the gas and so, a more complete analysis of the thermal instability would require a corresponding equation for the degree of ionization (Defouw 1970; Goldsmith 1970).

Thermal instability of a plasma in a uniform magnetic field has also been studied by Field (1965). It is obvious that for perturbations with wave vectors parallel to the field, the field does not affect the dynamics. If the wave vector is normal to the magnetic field, the magnetic field reduces the growth rate of the unstable modes.  Thus, a purely transverse magnetic field can prevent the thermal condensation (e.g., Oran, Mariska \& Boris 1982; Loewenstein 1990; Piontek \& Ostriker 2004). Hennebelle \& P$\acute{\rm e}$rault  (2000) studied magneto-thermal instability analytically and numerically in the general case corresponding to the neutral interstellar medium. David \& Bregman (1989) studied the effects of heat conduction and magnetic fields on the growth rate of thermal instabilities in cooling flows. Recently, Fukue \& Kamaya  (2007) revisited the effect of the ion-neutral friction of the two fluid on the growth of the thermal instability.

It is also well-known that dust particles constitute an ubiquitous and important component of many astrophysical plasmas including interstellar medium, stellar and planetary atmospheres, planetary nebulae and giant HII regions (e.g., Draine 2003). Dust particles can alter dynamical and thermodynamical properties of the plasma by transforming part of thermal energy into radiation (e.g., Kopp \& Shchekinov 2007; Bora 2004; Ib$\rm \acute{a}\tilde{n}$ez \& Shchekinov  2002). Birk \& Wiechen (2001) have considered the radiative condensation modes in a dense dusty plasma, where electrons are completely replaced by massive dust particles. However, in their analysis dust particles were considered only from the point of view of electrostatic interactions, while cooling processes were treated as independent of the presence of dust. Shukla \& Sandberg (2003)  studied radiation-condensation instability in a self-gravitating dusty astrophysical plasma. Kopp \& Shchekinov (2007) showed that positively  charged dust particles strongly destabilize perturbations.

In our model, neutrals are the thermodynamically active component and providing conditions for radiative cooling of the system, while ions, electrons and dust particles are the passive components. We show that when charged dust particles do not contribute to the cooling of the system, thermal stability properties of the system can modify. In the next section, we present general formulation of a multifluid system including charged dust particles. Analysis of the unstable modes are given in section 3. We conclude by a summary of the results in the final section.

\section{General Formulation}
\label{sec:2}

\subsection{Basic multifluid equations}

For the sake of simplicity we will describe in what follows the dust component as an ensemble of particles of equal masses, $m_{\rm d}$, and equal  charges, i.e. $Ze$. Our basic equations and the main assumptions are similar to Pandey \& Vladimirov (2007).
We take account of the different bulk velocities and densities of the neutral, electrons, ions and charged dust particles in the system. The continuity equation is
\begin{equation}
\frac{\partial\rho_{j}}{\partial t}+\nabla.(\rho_{j} {\bf v}_{j})=0,\label{eq:conj}
\end{equation}
where $\rho_{j}$ and ${\bf v}_{j}$ are the velocity of the various plasma components and the neutrals, respectively.

The momentum equations are
\begin{equation}
0=-q_{j}n_{j}({\bf E'}+ \frac{{\bf v}_{j}\times {\bf B}}{c})-\rho_{j}\nu_{jn}{\bf v}_{j},\label{eq:momj}
\end{equation}
\begin{equation}
\rho_{n}(\frac{\partial {\bf v}_{n}}{\partial t}+ {\bf v}_{n}.\nabla {\bf v}_{n})=-\nabla P + \sum_{e,i,d} \rho_{j}\nu_{jn}{\bf v}_{j}.\label{eq:momn}
\end{equation}
Note that velocities ${\bf v}_{j}$ are written in the neutral frame and ${\bf E'}={\bf E}+ {\bf v}_{n} \times {\bf B}/c$ is the electric field in the  neutral frame. $j$ stands for electrons ($q_{e}=-e$), ions ($q_{i}=e$) and dust ($q_{d}=Ze$), where $Z$ is the number of charge on the grain. The other physical variables have their usual meanings.

Also, the collision frequencies is (Draine et al. 1983; Draine 2003)
\begin{equation}
\nu_{jn}=\frac{<\sigma v>_{jn}}{m_{j}+m_{n}}\rho_{n},\label{eq:nu}
\end{equation}
where $<\sigma v>_{jn}$ is the rate coefficient for the momentum transfer by the collision of the $j^{\rm th}$ particle with the neutrals:
\begin{equation}
<\sigma v>_{in}=1.9\times 10^{-9} {\rm cm^{3} s^{-1}},
\end{equation}
\begin{equation}
<\sigma v>_{en}=4.5\times 10^{-9} (\frac{T}{30 \rm K})^{\frac{1}{2}} {\rm cm^{3} s^{-1}},
\end{equation}
and for small grains, we have $<\sigma v>_{dn} \approx <\sigma v>_{in}$, but for grains ranging between a few Angstrom to a few microns (Nakano \& Umebayashi 1986)
\begin{displaymath}
<\sigma v>_{dn}=2.8\times 10^{-5} (\frac{T}{30 \rm K})^{\frac{1}{2}}
\end{displaymath}
\begin{equation}
\times (\frac{a}{10^{-5} \rm cm})^{2} {\rm cm^{3} s^{-1}},\label{eq:dn}
\end{equation}
where $a$ is the grain radius. In our calculation, for the ion mass and mean neutral mass we adopt $m_{i}=30 m_{p}$ and $m_{n}=2.33 m_{p}$, where $m_{p}=1.67 \times 10^{-24} \rm g$ is the proton mass.

Defining the mass density of the bulk fluid and the bulk velocity as $\rho \approx \rho_{n}$ and ${\bf u} \approx {\bf v}_{n}$, equations (\ref{eq:conj}), (\ref{eq:momj}) and (\ref{eq:momn}) give the continuity and the momentum equations for the bulk fluid as
\begin{equation}
\frac{\partial \rho}{\partial t} + \nabla . (\rho {\bf u}) =0,\label{eq:main1}
\end{equation}
\begin{equation}
\rho (\frac{\partial {\bf u}}{\partial t}+ {\bf u}. \nabla {\bf u})=
- \nabla P + \frac{{\bf J}\times {\bf B}}{c}.\label{eq:main2}
\end{equation}

The next simplifying assumption is that electrons and ions are assumed well coupled to the magnetic field which implies $\beta_{e} \gg \beta_{i} \gg 1$, where $\beta_{j}=\omega_{cj}/\nu_{jn}$ is the ratio of cyclotron $\omega_{cj}=q_{j}B/m_{j}c$ to the collision frequencies. Based on this assumption and using quasi-neutrality condition, we have (Pandey \& Vladimirov 2007; Ciolek \& Mouschovias 1993)
\begin{equation}
{\bf v}_{e}=-\frac{1+\Theta}{Zen_{d}}{\bf J},
\end{equation}
where
\begin{equation}
\Theta=[1+\frac{\nu_{nd}}{\nu_{ni}}]\beta_{d}^{2}.\label{eq:Theta}
\end{equation}
Note that equation (\ref{eq:Theta}) is valid only for the radial component. General expression for $\Theta$ contains Hall and Pedersen conductivities (e.g., Ciolek \& Mouschovias 1993). Thus, the induction equation can be written as (Pandey \& Vladimirov 2007)
\begin{equation}
\frac{\partial {\bf B}}{\partial t} = \nabla \times [({\bf u}\times{\bf B}) -\frac{1+\Theta}{Zen_{d}} {\bf J}\times {\bf B} ]\label{eq:induc}
\end{equation}

We assume that dust particles do not contribute in cooling of the system. One should stress, however, that at temperature $3\times (10^7 - 10^8) \rm K$ dust particles are efficiently destroyed in collisions with the ions and electrons (Draine \& Salpeter 1979), and therefore only in the temperature range between $10^{6} \rm K$ and $3\times 10^7 \rm K$ does dust contribute sufficiently to the net cooling. So, we write energy equation as
\begin{equation}\label{eq:ENERGY}
\frac{1}{\gamma -1}\frac{dp}{dt}-\frac{\gamma}{\gamma -1}\frac{p}{\rho}\frac{d\rho}{dt}+\rho \Omega - \nabla . [K_{\parallel}\nabla_{\parallel} T + K_{\perp} \nabla_{\perp} T] =0,
\end{equation}
where the total time derivative is $d/dt = \partial / \partial t + {\bf u}.\nabla$, and $\gamma$ is the ratio of specific heats. Also, $\Omega$ represents the energy losses minus energy gains per unit mass. The coefficient of thermal conductivity $K$ has the values $K_{\parallel}$ and $K_{\perp}$ in directions parallel to and perpendicular to the magnetic field $\bf B$.

The net cooling function $\Omega$ is written depending on the physical conditions of the systems. For example, we are interested in the structure formation in a typical H I
dusty region  with C II cooling. We assume that $T_{0}=100$ K and $K=4.9\times 10^{3}$ ergs cm$^{-1}$ s$^{-1}$ K$^{-1}$ (Ulmschneider 1970). Then,
C II cooling is dominant when the temperature is $100$ K, according to Wolfire et al (1995). Thus,
\begin{displaymath}
\Omega = \frac{1}{m_{\rm H}} 2.54 \times 10^{-14} A_{\rm C} f_{\rm C {\rm II}} (\gamma^{H^{0}} n_{H^{0}}+\gamma^{e} n_{e})
\end{displaymath}
\begin{equation}
\times \exp(-92/T) {\rm erg }{\rm s}^{-1} {\rm g}^{-1},\label{eq:CoolingHI}
\end{equation}
where the fluid consists of H atoms with mass $m_{\rm H}$. Also, $n_{\rm X}$ is the density of element $X$, the constant $f_{\rm C {\rm II}}$ is the fraction of C I in C II, and $A_{\rm C}=n_{\rm C}/(n_{\rm H^{+}}+n_{H^{0}})= 3\times 10^{-4}$.
 The collisional de-excitation rate coefficients of C II with neutral hydrogen and electrons are represented by $\gamma^{H^{0}}$ and $\gamma^{e}$, respectively. According to Wolfire et al (1995),
\begin{equation}
\gamma^{H^{0}}=8.86\times 10^{-10} {\rm cm}^{3} {\rm s}^{-1},
\end{equation}
\begin{displaymath}
\gamma^{e}=2.1 \times 10^{-7} (T/100)^{-0.5} [ 1.80 + 0.484
\end{displaymath}
\begin{equation}
\times (T/ 10^{4} )+ 4.01 (T/ 10^{4} )^{2}-3.39 (T/ 10^{4})^{3}  ].
\end{equation}

In writing energy equation, we implicitly assumed that the equilibrium of the abundances of charged dust particles is established. Not only the ionization level, but the electrical charge of the grains and eventually, cooling function, may change due to the different mechanisms of interactions between the species.  However,  as long as the time scales of the ionization, recombination and the grain charging are much shorter than the cooling time scale, it is safe to neglect such complicated processes. In fact, The basic set of equation (\ref{eq:conj})-(\ref{eq:ENERGY}) is structurally similar to the single fluid ideal Hall MHD equations. This structural similarity occurs because of the collision which glues the weakly ionized medium together as a single fluid (e.g., Pandey and Wardle 2008) and therefore, the direct effect of recombination (which occurs on the collisional time scale) on the cooling function can be ignored (see also, Draine and Sutin 1987). We are interested to study thermal instability in a typical dusty H ${\rm I}$ region. For such a system it has been shown by Stiele, Lesch \& Heitsch (2006) and  Ciolek \& Mouschovias (1993) that the cooling time scale is larger than the other important time scales.

Finally, we can write equation of state as
\begin{equation}\label{eq:STATE}
p=\frac{R}{\mu} \rho T,
\end{equation}
where $R$ is the gas constant and $\mu$ represents the molecular weight.

Equations (\ref{eq:main1}), (\ref{eq:main2}), (\ref{eq:induc}), (\ref{eq:ENERGY}) and (\ref{eq:STATE}) along with the equation
\begin{equation}
\nabla . {\bf B}=0,
\end{equation}
are our basic equations for magneto-thermal instability in a partially ionized medium with charged dust particles.

\subsection{Linear perturbations}
The set of the equations can be linearized in the absence of gradients in the initial state. We assume that the initial equilibrium state are  characterized by the values $\rho_0$,  $P_{0}$, $T_{0}$ and ${\bf B}_{0}$, and the velocity is assumed to be zero in the nonperturbed initial equilibrium state of the fluid and $\Omega(\rho_0, T_0)=0$.

We assume perturbations of the form
\begin{equation}\label{perturb}
\Pi(\textbf{r},t) = \Pi_{1} \exp
 (\omega t + i \textbf{k} \cdot \textbf{r}),
\end{equation}
where $\Pi_{1}$ is the amplitude of the perturbations, $\omega$ is the growth rate of the perturbations and ${\bf k}$ is the wavenumber of the perturbations. A condensation mode occurs when $\omega$ is real and positive. If $\omega$ is real and negative, the perturbations damp. But when $\omega$ is a complex number, the system is oscillatory growing or damping depending on the sign of $\Re (\omega)$. The system is called overstable, if we have $\Re(\omega) > 0$. We mainly interested in the possible effects of the charged dust particles on the condensation modes.

Then, the linear equations are
\begin{equation}\label{masslin}
\omega \rho_1 + i \rho_0 \textbf{k} \cdot \textbf{v}_1=0,
\end{equation}
\begin{equation}\label{momenlin}
\omega \rho_0 \textbf{v}_1+ i \textbf{k}  P_1+ i
(\textbf{B}_0\cdot\textbf{B}_1) \frac{\textbf{k}}{4\pi}-
i(\textbf{k}\cdot\textbf{B}_0)\frac{\textbf{B}_1} {4\pi}=0,
\end{equation}
\begin{displaymath}\label{energlin}
\frac{\omega}{\gamma-1}P_1-\frac{\omega\gamma P_0}{(\gamma-1)\rho_0}\rho_1+
\rho_0 \Omega_\rho \rho_1 +\rho_0 \Omega_T T_1
\end{displaymath}
\begin{equation}
+(K_{\parallel} k_{\parallel}^{2} + K_{\perp} k_{\perp}^{2})T_{1}=0,
\end{equation}
\begin{eqnarray}\label{maglin}
\nonumber \omega \textbf{B}_1 +i \textbf{B}_0 (\textbf{k} \cdot
\textbf{v}_1) - i (\textbf{k} \cdot\textbf{B}_0) \textbf{v}_1 -
\\
-\frac{c}{4\pi}\frac{1+\Theta}{Ze n_{\rm d}} (\textbf{k} \cdot \textbf{B}_0) (\textbf{k} \times
\textbf{B}_1)=0,
\end{eqnarray}
\begin{equation}\label{idealin}
  \frac{P_1}{P_0} - \frac{\rho_1}{\rho_0} - \frac{T_1}{T_0}=0.
\end{equation}
Note that the derivative $\Omega_{\rho}=(\partial\Omega/\partial\rho)_{T}$ and $\Omega_{T}=(\partial\Omega/\partial T)_{\rho}$ are evaluated for the equilibrium state.

\begin{figure}
\epsfig{figure=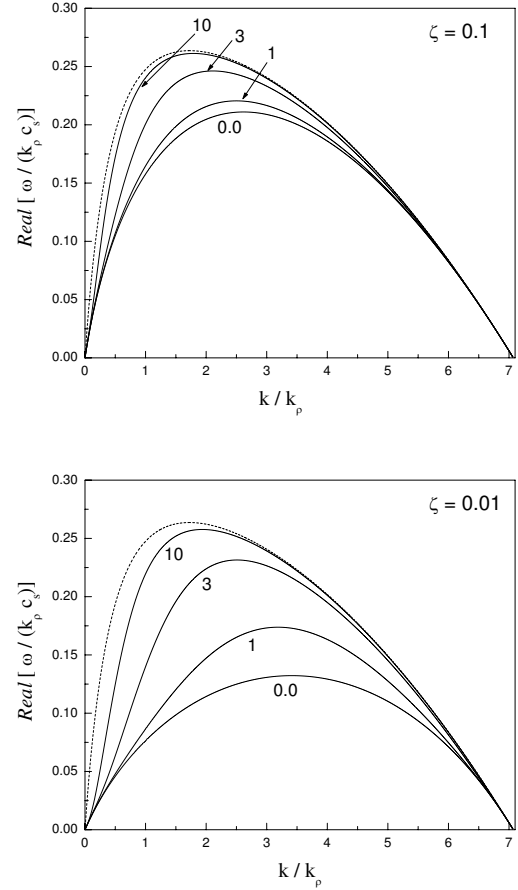,angle=0,width=10.0cm}
\caption{Growth rate of magneto-thermal condensation mode versus wavenumber of the perturbations when $\alpha=1$, $\gamma=5/3$, $\sigma_{\rm T}/\sigma_{\rho} = 1/2$ and $\sigma_{\rho} \sigma_{K} = 0.01$. Dashed line represents growth rate in non-magnetized case and without charged dust particles. Each curve is labeled by the ratio $\sqrt{\Lambda} = k_{\rho}c_{\rm s}/\omega_{\rm mcd}$.}
\label{fig:f1}
\end{figure}

\subsection{Dispersion relation}
We introduce the coordinate system $\textbf{\textit{e}}_x$,
$\textbf{\textit{e}}_y$, and $\textbf{\textit{e}}_z$ specified by
\begin{equation}\label{coord}
\textbf{\textit{e}}_z=\frac{\textbf{B}_0}{B_0}\quad,\quad\textbf{\textit{e}}_y=\frac{\textbf{B}_0\times\textbf{k}}
{|\textbf{B}_0\times\textbf{k}|}\quad,\quad\textbf{\textit{e}}_x=\textbf{\textit{e}}_y\times\textbf{\textit{e}}_z.
\end{equation}
Also, we introduce the following wavenumbers
\begin{displaymath}
k_{\rho}=\mu (\gamma -1) \rho_{0}\Omega_{\rho} (Rc_{\rm s} T_{0})^{-1}, k_{T} = \mu (\gamma -1 ) \Omega_{T} (Rc_{\rm s})^{-1},
\end{displaymath}
\begin{displaymath}
k_{K_{\parallel}} = [\mu (\gamma -1) K_{\parallel}]^{-1} (Rc_{\rm s}\rho_{0}),
\end{displaymath}
\begin{equation}
k_{K_{\perp}} = [\mu (\gamma -1) K_{\perp}]^{-1} (Rc_{\rm s}\rho_{0}).
\end{equation}
Then, we can write the dispersion equation using the following non-dimensional quantities,
\begin{displaymath}
\Gamma = \frac{\omega}{k c_{\rm s}}, \sigma_{\rho} = \frac{k_{\rho}}{k}, \sigma_{T}=\frac{k_{T}}{k}, \sigma_{K_{\parallel}}=\frac{k}{k_{K_{\parallel}}}, \sigma_{K_{\perp}}=\frac{k}{k_{K_{\perp}}}.
\end{displaymath}

Thus, the characteristic equation becomes
\begin{displaymath}
\Gamma^{7}+P_6 \Gamma^6 + P_5 \Gamma^5 + P_4 \Gamma^4 + P_3 \Gamma^3 + P_2 \Gamma^2
\end{displaymath}
\begin{equation}\label{eq:dis}
+ P_1 \Gamma + P_0=0,
\end{equation}
where the coefficients are
\begin{displaymath}
P_0 = \alpha^2 \zeta^2 \gamma^{-1}(\sigma_{\rm T}+\sigma_{\rm K}-\sigma_{\rho}),
\end{displaymath}
\begin{displaymath}
P_1 = \zeta^2 \alpha^2,
\end{displaymath}
\begin{displaymath}
P_2 = \alpha^2 \zeta (\sigma_{\rm T}+\sigma_{\rm K}) - \alpha \zeta \gamma^{-1} (\sigma_{\rho}-\sigma_{\rm T}-\sigma_{\rm K})
\end{displaymath}
\begin{displaymath}
\times [2+ \alpha \Lambda (k/k_{\rho})^{2} ],
\end{displaymath}
\begin{displaymath}
P_3 = \alpha \zeta [2+\alpha + \alpha \Lambda (k/k_{\rho})^{2} ],
\end{displaymath}
\begin{displaymath}
P_4 = \alpha (\sigma_{\rm T}+\sigma_{\rm K}) [1+ \zeta +\alpha\zeta \Lambda(k/k_{\rho})^{2} ]
\end{displaymath}
\begin{displaymath}
-\gamma^{-1} (\sigma_{\rho}-\sigma_{\rm T}-\sigma_{\rm K}),
\end{displaymath}
\begin{displaymath}
P_5 = 1 + \alpha + \alpha \zeta [1+ \alpha \Lambda (k/k_{\rho})^{2} ],
\end{displaymath}
\begin{equation}
P_{6}=\sigma_{\rm T}+\sigma_{\rm K},
\end{equation}
and $\zeta = \cos^{2}\theta$ and $\theta$ is the angle between ${\bf B}_{0}$ and ${\bf k}$ and $\alpha = (v_{\rm A}/c_{\rm s})^2$ where $v_{\rm A}$ is the Alfven velocity. Also, we have $\sigma_{K}=\sigma_{K_{\parallel}} \zeta + \sigma_{K_{\perp}} (1-\zeta)$. However, we note that thermal conductivity in a plasma is suppressed perpendicular to the magnetic field if the electron gyroradius is much smaller than the collisional mean free path (Spitzer 1962), as is generally true under interstellar conditions. Also, we have $\Lambda = (k_{\rho}c_{\rm s}/\omega_{\rm mcd})^{2}$, where $\omega_{\rm mcd}=(\rho_{\rm d}/\rho_{0})(1/1+\Theta)(ZeB/m_{\rm d}c)$ is  the  modified dust-cyclotron frequency and $(k_{\rho} c_{\rm s})^{-1}$ represents the cooling time-scale. Thus, parameter $\sqrt{\Lambda}$ is ratio of the cooling rate and the  modified dust-cyclotron frequency. We see that possible effects of the charged dust particles on the unstable modes of equation (\ref{eq:dis}) appears through parameter $\Lambda$. If we set $\Lambda = 0$, our dispersion equation (\ref{eq:dis}) reduces to the standard characteristic equation of the magneto-thermal instability, but without dust particles (Field 1965). In this case, dispersion equation (\ref{eq:dis}) becomes
\begin{displaymath}
(\Gamma^2+\alpha \zeta)\{ \Gamma^5 + (\sigma_{\rm T}+\sigma_{\rm K}) \Gamma^4 + (1+\alpha)\Gamma^3+[ \alpha (\sigma_{\rm T}+\sigma_{\rm K})
\end{displaymath}
\begin{equation}
+(\sigma_{\rm K}+\sigma_{\rm T}-\sigma_{\rm\rho})/\gamma ] \Gamma^2 + \alpha \zeta \Gamma + \alpha \zeta (\sigma_{\rm T}+\sigma_{\rm K}-\sigma_{\rho})/\gamma \}=0.
\end{equation}
The expression inside the above parenthesis describes stable modes. But fifth order polynomial that has been obtained by many authors (e.g., Field 1965; Stiele et al 2006) gives us magnetothermal condensation modes without charged dust particles.

\section{Analysis}

\begin{figure}
\epsfig{figure=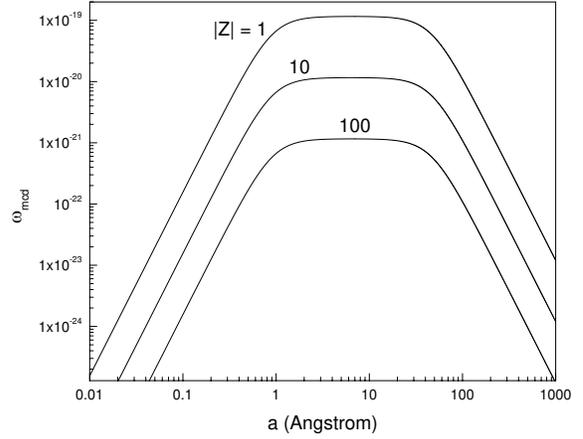,angle=0,width=10.0cm}
\caption{Modified dust-cyclotron frequency $\omega_{\rm mcd}$ versus radius of the dust particles (in Angstrom) with various electrical charge. We assume the level of the ionization is $\rho_{\rm i}/\rho_{0} =10^{-6}$ and the magnetic field strength is $B_{0}=10^{-6}$ G. Also, $\rho_{0} /m_{\rm p}=100$ cm$^{-3}$ and $\rho_{d}/\rho_{0} =0.01$. Each curve is labeled by $|Z|$.}
\label{fig:f2}
\end{figure}

Now, we can study the effects of charged dust particles on the condensation modes in typical plasma containing charged dust particles. Equation (\ref{eq:dis}) is of odd degree in $\Gamma$ and must therefore admit at least one positive real root for non-perpendicular perturbations (i.e., $\zeta\neq 0$) if the last term $P_{0}$ is negative. But when $\zeta$ becomes zero (i.e, wavevector is transverse to the magnetic field) the dispersion equation (\ref{eq:dis}) reduces to the classical case of the transverse perturbations (see Field 1965, Eq.  61). In this case, Hall like term (due to the existence of the charged dust particles) in equation (\ref{eq:induc}) does not contribute to the problem and the criterion of the instability becomes $P_{4}<0$ that is consistent to the  previous studies (e.g., Field 1965). Thus, we see that the condition for a positive real root (implying monotonic instability) for a transverse field is independent of the presence of charged dust particles. Note that in our analysis, we assumed that dust particles do not contribute to the net cooling of the system. The other roots may be complex conjugates of each other correspond to either damped waves or growing waves. If all roots of equation (\ref{eq:dis}) have negative real parts, then the system is stable magneto-thermally. Using Hurwitz analysis, we determined conditions under which all roots have negative real parts. We noticed that conditions of the stability are the same as magnetized case without dust particles. In other words, dust particles do not change criteria of the stability of the system.

In the absence of dust particles, the maximum effect of magnetic field occurs for perturbations perpendicular to the initial magnetic field, i.e $\zeta=0$. For such perturbations, dust particles do not change growth rate of the condensation modes because all terms including $\Lambda$ vanish when $\zeta=0$. So, nearly perpendicular perturbations are considered in our analysis.

We solve numerically the roots of dispersion equation (\ref{eq:dis}) for some values of the nondimensional parameters and taking $\gamma = 5/3$. We take the parameters $\alpha=1$, $\sigma_{\rm T}/\sigma_{\rho} = 1/2$ and $\sigma_{\rho} \sigma_{K} = 0.01$ for comparison to Field (1965). Figure \ref{fig:f1} shows growth rate of the condensation modes for nearly perpendicular perturbations, i.e. $\xi=0.1$ and $0.01$.  Each curve is labeled by corresponding ratio $k_{\rho}c_{\rm s}/\omega_{\rm mcd}=\sqrt{\Lambda}$. For comparison, growth rate of the non-magnetized condensation mode without dust particles is shown by dashed lines in Figure \ref{fig:f1}. We can see that growth rate in the magnetized case without dust particles (i.e. $\Lambda=0$) is lower than  the growth rate in the non-magnetized case. Charged dust particles enhance growth rate of the condensation mode and parameter $\Lambda$ has a vital role. When cooling rate is smaller than the modified dust-cyclotron frequency, enhancement of the growth rate due to the dust particles is negligible. But as the cooling rate becomes larger than the modified dust-cyclotron frequency, growth rate of the unstable mode significantly increases and tends to the non-magnetized growth rate when parameter $\Lambda$ is large.

\begin{figure}
\epsfig{figure=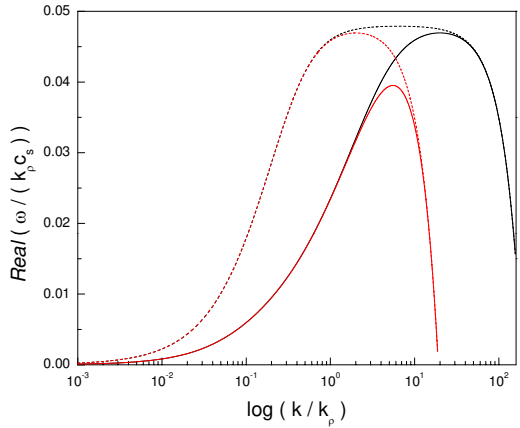,angle=0,width=11.0cm}
\caption{Growth rates of the magneto-thermal unstable mode in the typical dusty H ${\rm I}$ region, where $T=100$ K, $f_{\rm C{\rm II}}=0.01$,  $\rho_{0} = 100$ cm$^{-3}$, $\rho_{\rm i}/\rho = 10^{-6}$, $\rho_{\rm d}/\rho_{0}=0.01$ and $B_{0}=10^{-6}$. Also, we have $\alpha=0.048$, $\gamma=5/3$, $\sigma_{\rm T}/\sigma_{\rho} =0.92$. Black curves are corresponding to a case with $\sigma_{\rho} \sigma_{K} = 2.19\times 10^{-6}$, and for comparison, red curves are for $\sigma_{\rho} \sigma_{K} = 2.19\times 10^{-4}$. Solid curves are representing unstable modes without charged dust particles. Radius of dust particles is $a=1 \AA$ and the electrical charge is $|Z|=1$. It is evaluated $k_{\rho}c_{\rm s}=9.28\times 10^{-14}$ s and $\omega_{\rm mcd}=6.68\times 10^{-20}$ s.}
\label{fig:f3}
\end{figure}

Profile of the growth rate reaches to a maximum value for a wavenumber $k_{\rm max}$ which depends on the input parameters. Wavenumber $k_{\rm max}$ corresponding to the maximum growth rate increases because of the magnetic field in the absence of the dust particles. In other words, the most unstable mode occurs at a smaller wavelength because of the magnetic field. But wavenumber $k_{\rm max}$ decreases because of the charged dust particles. Note that growth rate of the perturbation parallel to  the initial magnetic field line does not modify because of the charged dust particles. But as perturbations tend to the perpendicular direction to the field line, the influence of charged dust particles on the growth rate becomes more significant.

Figure \ref{fig:f2} shows the modified dust-cyclotron frequency, $\omega_{\rm mcd}$, versus the grain radius $a$ for different electrical charge. We assumed  the number density of the neutral component and the magnetic field strength are about $100$ cm$^{-3}$ and about $10^{-6}$ G, respectively. Also, the ionization degree is assumed to be $\rho_{\rm i}/\rho_{0}=10^{-6}$. We see that the modified dust-cyclotron frequency decreases with the electrical charge. Based on the definition of the modified dust-cyclotron frequency, we see that $\omega_{\rm mcd} \propto |Z|/\Theta$. On the other hand, we have $\Theta \propto |Z|^2$. Thus, we can conclude $\omega_{\rm mcd} \propto |Z|^{-1}$. For increasing values of $|Z|$ the modified dust-cyclotron frequency $\omega_{\rm mcd}$ decreases which implies less magnetic coupling and larger $\Lambda$ for a fixed cooling rate. Since growth rate increases with $\Lambda$, we can conclude that a larger electrical charge $|Z|$ implies greater growth rate. We note that the dust grains can only respond to variations of the magnetic field with frequency less than $\omega_{\rm mcd}$.

Dependence of the dust-cyclotron frequency on the grain radius is interesting. While for the grains with radii between $1$ and $30$ Angstrom the modified dust-cyclotron frequency is constant more or less, we see that $\omega_{\rm mcd}$ increases with the grain radius for $a<1 {\rm \AA}$, and, $\omega_{\rm mcd}$ decreases with the grain radius for $a>30{\rm \AA}$. So, the level of magnetic coupling of the charged dust particles highly depends on the grain radius. Since $\omega_{\rm mcd}$ increases with $a$ for the grains with radii smaller than $1 {\rm \AA}$, the particles are more  well-coupled magnetically and moreover parameter $\Lambda$ decreases for a fixed cooling rate which implies smaller growth rate according to Figure \ref{fig:f1}. Minimum influence of the grains on the growth rate occurs for the grains with radii between $1$ and $30$ Angstrom,  for which dust-cyclotron frequency reaches to a maximum value. In this case, growth rate is independent of the grain radius more or less as long as the cooling rate is kept fixed. But as the grain radius becomes larger than $30 {\rm \AA}$, the frequency $\omega_{\rm mcd}$ decreases with radius $a$ and the dust particles become less coupled to the field lines. Then, parameter $\Lambda$ increases with the radius $a$ and the growth rate increases according to Figure \ref{fig:f1}.

Now, we can study magneto-thermal instability in a typical H ${\rm I}$ region with charged dust particles. Our input parameters are similar to Fukue \& Kamaya (2007) who analyzed thermal instability in a typical  H ${\rm I}$ by considering ions and neutrals, separately. But they neglected dust particles. Our analysis is restricted to weakly ionized case and the bulk density and velocity are determined by the neutrals. Figure \ref{fig:f3} shows the growth rates of the unstable modes when the ionization degree of a typical H ${\rm I}$ with temperature around $100$ K and the density $100$ cm$^{-3}$ is assumed to be $10^{-6}$. The strength of the magnetic field is $B_{0}=10^{-6}$ G. Having the net cooling function (\ref{eq:CoolingHI}), we can simply calculate that $k_{\rho}c_{\rm s}=9.28\times 10^{-14}$ s. Also, we have $f_{\rm C {\rm II}}=0.01$, $\rho_{\rm d}/\rho_{0}=0.01$, $\alpha=0.048$, $\gamma=5/3$, $\sigma_{\rm T}/\sigma_{\rho} =0.92$. Black curves are corresponding to a case with $\sigma_{\rho} \sigma_{K} = 2.19\times 10^{-6}$, and for comparison, red curves are for $\sigma_{\rho} \sigma_{K} = 2.19\times 10^{-4}$. Solid curves are representing unstable modes without charged dust particles. Radius of dust particles is $a=1 \AA$ and the electrical charge is $|Z|=1$. We can also simply calculate that $\omega_{\rm mcd}=6.68\times 10^{-20}$ s. Figure \ref{fig:f3} shows that the existence of the charged dust particles enhances the growth rate of the unstable modes. This enhancement is more significant as thermal conduction becomes more effective. Also, because of the charged dust particles the maximum wavenumber $k_{\rm max}$ shifts to the smaller values. It implies that in a typical H ${\rm I}$ region one may expect formation of larger structures due to the thermal instability with charged dust particles. Not only the structures are bigger, but they are forming faster comparing to a similar system without charged dust particles.

Although Fukue \& Kamaya (2007) did not give a simple physical explanation for their findings, we think stabilization of ions via drag force is understandable and it will help us to explain our results. First, we consider a two-fluid (i.e., ion and neutral) system without magnetic field. Direction of the exerted drag force on each charged species is opposite to the direction of its velocity. So, it acts as a restoring force in response to any perturbations.  Obviously, the effect is much stronger for ions because of their mean molecular weight comparing to the neutrals as was shown in Figure 1 of Fukue \& Kamaya (2007).   This effect is survived even in the presence of magnetic field lines. Because the magnetic tension vanishes for purely transverse perturbations and the magnetic pressure adds up to the thermal pressure which implies a more thermally stable system. In our analysis, as we discuss, the drag force is not significant comparing to the magnetic force at least for ions and electrons. So, response of the system to the perturbations is mainly determined by the distribution of the magnetic field lines. There is a drift velocity between charged dust particles and the neutrals. This fact is shown by a modified induction equation with an extra term similar to the Hall term. Therefore, we can expect dissipation of the magnetic field and reduction to the magnetic pressure due to the presence of charge dust particles. Charged dust particles are well tied to the magnetic field lines, i.e. $\Lambda \approx 0$.  When this parameter increases due to the reduction of $\omega_{\rm mcd}$   (for a fixed cooling rate), the dust particles are less coupled to the field lines comparing to a case with $\Lambda \approx 0$. Thus, the coefficient of the Hall term increases that is proportional to the inverse of  $\omega_{\rm mcd}$.

There are also more points before comparing our results to Fukue \& Kamaya (2007). In our approach that is based on Pandey \& Vladimirov (2007), ions and electrons are strongly tied to the magnetic field lines and the density of the charged species are neglected comparing to the bulk density of the system that is controlled by the neutrals. Also, bulk velocity of the system is determined by the neutral particles. But analysis of Fukue \& Kamaya (2007) is within ambipolar regime by considering the complete sets of equations for a two-fluid (i.e., ion and electron) system. In other words, although we are studying a multifluid system (including charged dust particles), our main equations are still single fluid equations. More importantly, our equations are valid as long as ions and electrons are coupled to the magnetic field lines. It implies that the drag force is negligible comparing to the electromagnetic force. For this reason collision does not seem to play a significant role in dynamics of ions and electrons. On the other hand, behavior of charged dust particles to the perturbations is understood by the modified cyclotron frequency.
There is also another important point: Our result does not contradict Fukue \& Kamaya (2007) regarding to the reduction of the growth rate due to collision.  Actually, our results are supportive to them. Let's explain how. First of all, Fukue \& Kamaya (2007) mentioned (in the abstract of their paper) that the instability is suppressed via the friction. But it is true for ions as they clearly discussed in section 4.2.3 of Fukue \& Kamaya (2007). So, significant reduction in growth rate is actually for ions (not neutrals). But in our analysis, we are employing a single fluid approach and what we observe as (in)stability is for neutrals.

\section{Conclusion}

We studied magneto-thermal instability in the presence of charged dust particles with little mobility. Although dust particles may contribute to the net cooling function, their contribution can be neglected for some ranges of temperature (e.g., Draine \& Salpeter 1979). The relative drift between the plasma and the charged dust grains lead to the Hall term in the resulting induction equation. It is know that this feature is a generic property of the multifluid plasmas (e.g., Wardle 1999; Pandey \& Vladimirov  2007). The inclusion of charged dust particles does not alter the criterion for magneto-thermal stability, but the growth rate of the unstable modes are significantly modified.

We showed that the growth rate of the condensation mode depends on the ratio of the cooling rate and the modified dust-cyclotron frequency. Growth rate increases with the ratio. Thus, as charged dust particles become less well-coupled to the field lines, the growth rate becomes larger and the system is more stable to the perturbations as long as the cooling rate does not change.  Also, the growth rate increases with the electrical charge. While the growth rate increases with the grain radius when they are larger than $30$ Angstrom, we see opposite behavior when grains are smaller than $1$ Angstrom. Magneto-thermal condensation mode is independent of  the grain radius, when they are between $1$ and $30$ Angstrom. Note that these critical sizes may change for the another set of input parameters, but the profile of $\omega_{\rm mcd}$ is something similar to Figure \ref{fig:f2} for the other input parameters.

Our analysis shows that even a small amount of the charged dust particles may modify magneto-thermal condensation modes. However, a more detailed analysis is needed which should also take into account the dynamics of the dust and neutral particles.

\section*{Acknowledgments}

I am grateful  to the referee for very useful comments and  suggestions to improve the paper.

{}

\end{document}